# Johari-Goldstein relaxation in glass-electrets

P. Lunkenheimer,[1,*] F. Humann,[1] D. Reuter,[1] K. Geirhos,[1] A. Loidl,[1] and G. P. Johari[2]
[1]*Experimental Physics V, Center for Electronic Correlations and Magnetism, University of Augsburg, 86159 Augsburg, Germany*
[2]*Department of Materials Science and Engineering, McMaster University, Hamilton, ON L8S 4L7, Canada*

We investigate the dielectric response in the glass-electret state of two dipolar glass-forming materials. This unusual polar glassy state of matter is produced when a dipolar liquid is supercooled under the influence of a high electric dc field, which leads to partial orientational order of the molecules carrying a dipole moment. Investigation of the prepared glass-electrets by using low-field dielectric spectroscopy reveals a clear modification of their dielectric response in the regime of the Johari-Goldstein $\beta$-relaxation, pointing to a small but significant increase of its relaxation strength compared to the normal glass. We discuss the implications of this finding for the still controversial microscopic interpretation of the Johari-Goldstein relaxation, an inherent property of glassy matter.

Dielectric spectroscopy is an essential experimental method for investigating the still not fully understood glass transition and the glassy state of matter. Usually, it is used to study dipolar glass formers whose response to relatively small ac fields is detected, ensuring that the sample properties are not altered by the applied electrical field. Relying on the fluctuation-dissipation theorem, such conventional, low-field dielectric spectroscopy enables one to obtain information on molecular dynamics of polarization fluctuations. In recent years a markedly different approach has come into the focus of interest: In the so-called nonlinear dielectric experiments, a glass former is subjected to high dc electrical fields of up to several hundreds kV/cm, which lead to a field-induced change of the material properties, e.g., the equilibrium permittivity or a change of the slope in the polarization-versus-field plot at high fields. Such an experiment reveals the nonlinear dielectric properties of a material, e.g., the higher harmonics of the dielectric susceptibility (for an overview, see Refs. [1,2] and the collection of articles in Ref. [3]). Significant conclusions about dynamic and static heterogeneities [4,5,6], entropy [7,8,9], and cooperativity of molecular motions in glass-forming matter have been drawn from such studies [10,11,12,13].

In such experiments, usually a high ac or dc electric field is applied *during* the dielectric measurement. We propose a different approach, which is to separate the application of the high dc field from the actual measurement: In a first step, a material may be subjected to a high dc field for a certain time, thereby transferring it into a different state. In the second step, its dielectric properties may be measured conventionally, i.e., with low ac field. In glass-forming matter, such experiments can only work close to or below the glass-transition temperature $T_g$, because, at higher temperatures, the field-induced modification of the material would quickly decay. When first applying a high *ac* field, this corresponds to the so-called dielectric hole-burning experiments, which have helped to prove that there is a heterogeneity-induced distribution of relaxation times in glass-forming matter [14]. That a high *dc* field can also alter the glass state of a dipolar material is well known from the so-called thermally-stimulated depolarization-current (TSDC) measurements. There, first a strong dc field is applied during supercooling from the liquid deep into the solid glass. This freezes the dipolar orientations into an (at least partially) orientationally ordered state with permanent electrical polarization (see upper inset of Fig. 2 for a schematic illustration), which persists even after removal of the electric field. The sample is then heated without the field, leading to the recovery of orientational disorder above $T_g$ which triggers the detection of a small pyrocurrent. In contrast, in the present work we investigate the dielectric properties in the induced orientationally-ordered state below $T_g$.

In some respects, the field-induced permanent dipolar order, generated in such or similar experiments, resembles the residual magnetization caused by spin order as found in permanent magnets. Thus, in analogy to magnets, samples with such residual dipolar order were termed "electrets" [15,16] and, for those that are based on glassforming materials, the term "glass-electret" was coined [17]. A very common application of various types of electrets are electret microphones. Recently, especially the glass-electret state of matter has found renewed interest due to its promising pharmaceutical application [17]: A high solubility of a drug is essential for many medical applications and it is well known that in the amorphous state it can be significantly higher than in the crystalline form. Interestingly, in Ref. [17] it was shown that the solubility can be even further enhanced in the glass-electret state, which was rationalized by its higher free energy.

In the present work, we investigate the molecular dynamics in this unusual, partially ordered form of glassy matter. As this state is stable only below $T_g$, the primary molecular motions (termed "$\alpha$ relaxation") that mirror the strong increase of viscosity when approaching the glass transition, are essentially frozen and cannot be investigated in detail. However, it is well known that significantly faster secondary relaxation processes exist in nearly all classes of glassy matter, which can also be observed below $T_g$. Here we especially focus on the Johari-Goldstein (JG) relaxation, which was demonstrated to be an inherent property of the glassy matter [18,19] and proposed to be a precursor of the $\alpha$ relaxation [20,21,22]. However, in spite of its high relevance for glassy dynamics and for the glass transition in general and despite a long history of research devoted to the JG relaxation, there is still a lack of consensus concerning its microscopic origin. The different proposed microscopic explanations of this phenomenon essentially can be divided into two groups: (i) that *all* molecules participate in the JG relaxation [23,24,25,26] and (ii) only a fraction of the total number of molecules participate in this process [19,27]. In the first case, in addition to the $\alpha$ relaxation, the molecules are typically assumed to also perform faster motions with smaller displacements than the $\alpha$ relaxation (e.g., small-angle reorientations, whereas the $\alpha$ relaxation is a full rotation). (However, recently it has been argued [22] that the view that all molecules participate in the JG process is inconsistent with the general understanding of density and entropy fluctuations.) In the second case, the JG relaxation is usually supposed to arise from molecules located in so-called "islands of mobility", where these molecules have higher mobility, e.g., due to a reduced density within the islands [19,22,27].

It is well known that a modification of the glass state, e.g., by very rapid cooling or by applying pressure during cooling can strongly affect the properties of the JG relaxation, e.g., the relaxation time or amplitude [27,28,29,30,31,32]. Are there similar effects for the so-far only rarely investigated glass-electret state? Can we achieve a better understanding of the microscopic origin of the JG relaxation by exploring its behavior in this unusual glassy state of matter? In the present work, we try to answer these questions.

For our study, we have chosen two glass-forming epoxy compounds, diglycidyl ether of bisphenol-A (DGEBA; also known by its trade name EPON828) and triphenylolmethane triglycidyl ether (TPMTGE). Both are very good glass formers and exhibit two secondary relaxations, which are well separated from the $\alpha$ relaxation in the dielectric spectra below $T_g$ [33,34,35,36]. Based on model considerations and empirical findings [37,38], in Refs. [20,39] the slower of these secondary relaxations was identified as "genuine" JG relaxation. The faster relaxation, termed $\gamma$ relaxation, was suggested to be due to intramolecular motions.

DGEBA and TPMTGE were purchased from Sigma-Aldrich. The liquid sample material was put between two lapped and highly polished stainless-steel plates, with glass-fiber spacers ensuring a plate distance of 30 µm. For this purpose, TPMTGE had to be slightly heated above its melting point of 321 K while DGEBA is a liquid with rather high viscosity already at room temperature. For the dielectric measurements, both in the glass-electret and normal-glass state, a frequency-response analyzer (Novocontrol Alpha-A analyzer) was used. For temperature variation, the sample was put into a $N_2$-gas flow cryostat.

The glass-electret was prepared by cooling the sample at a rate of 0.4 K/min from a temperature $T_{high} > T_g$ to a temperature $T_{low} < T_g$ with an applied dc field of 167 kV/cm (DGEBA: $T_g \approx 255$ K [35], $T_{high} = 280$ K, $T_{low} = 150$ K; TPMTGE: $T_g \approx 287$ K [36], $T_{high} = 291$ K, $T_{low} = 160$ K). After switching off the dc field, conventional dielectric spectroscopy with a moderate ac field of 0.33 kV/cm was performed during heating the sample up to $T_{high}$ with 0.4 K/min. This cooling/heating cycle was repeated without applying a dc field under cooling, which allowed the detection of the dielectric response in the normal-glass state for comparison. By using a TSDC study of DGEBA, we have demonstrated that already a field of 133 kV/cm is sufficient to generate significant polarization in the glass state (Fig. S1 in the Supplemental Material [40]; see also [41]).

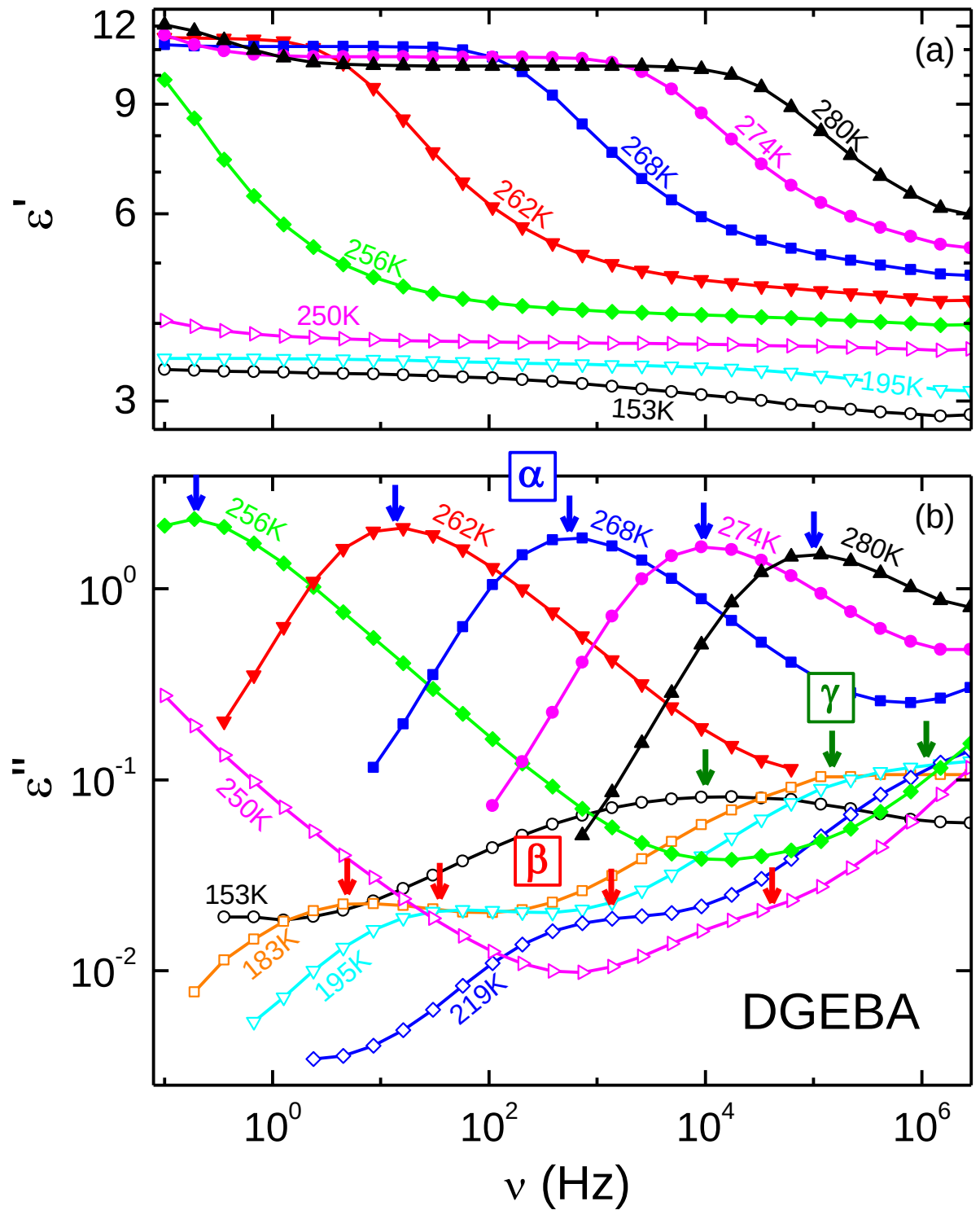

FIG. 1. Spectra of the dielectric constant $\varepsilon'$ (a) and loss $\varepsilon''$ (b) of DGEBA at selected temperatures (to keep the figure readable, for $\varepsilon'$ less low-temperature curves are shown than for $\varepsilon''$). The lines connect the data points. The arrows indicate the peak frequencies of the three observed relaxation processes.

In addition, conventional dielectric measurements in a broader temperature range were performed to provide an overview of the dielectric behavior of these materials. The results for DGEBA are shown in Fig. 1. In accord with previous investigations [33,34,35], the $\alpha$ relaxation is revealed by the dominating step in the dielectric constant $\varepsilon'(\nu)$ [Fig. 1(a)] and by the peak in the loss $\varepsilon''(\nu)$ [Fig. 1(b)] for the five highest shown temperatures. While in $\varepsilon'(\nu)$, at lower temperatures, a second, smeared out step with much lower amplitude is observed, the loss spectra disclose the existence of two secondary processes in this region. For TPMTGE, qualitatively similar behavior is found (see Fig. S2 in Supplemental Material [40] and Ref. [36]). As mentioned above, for both glass formers the slower secondary process was identified as the JG $\beta$-relaxation [20,39].

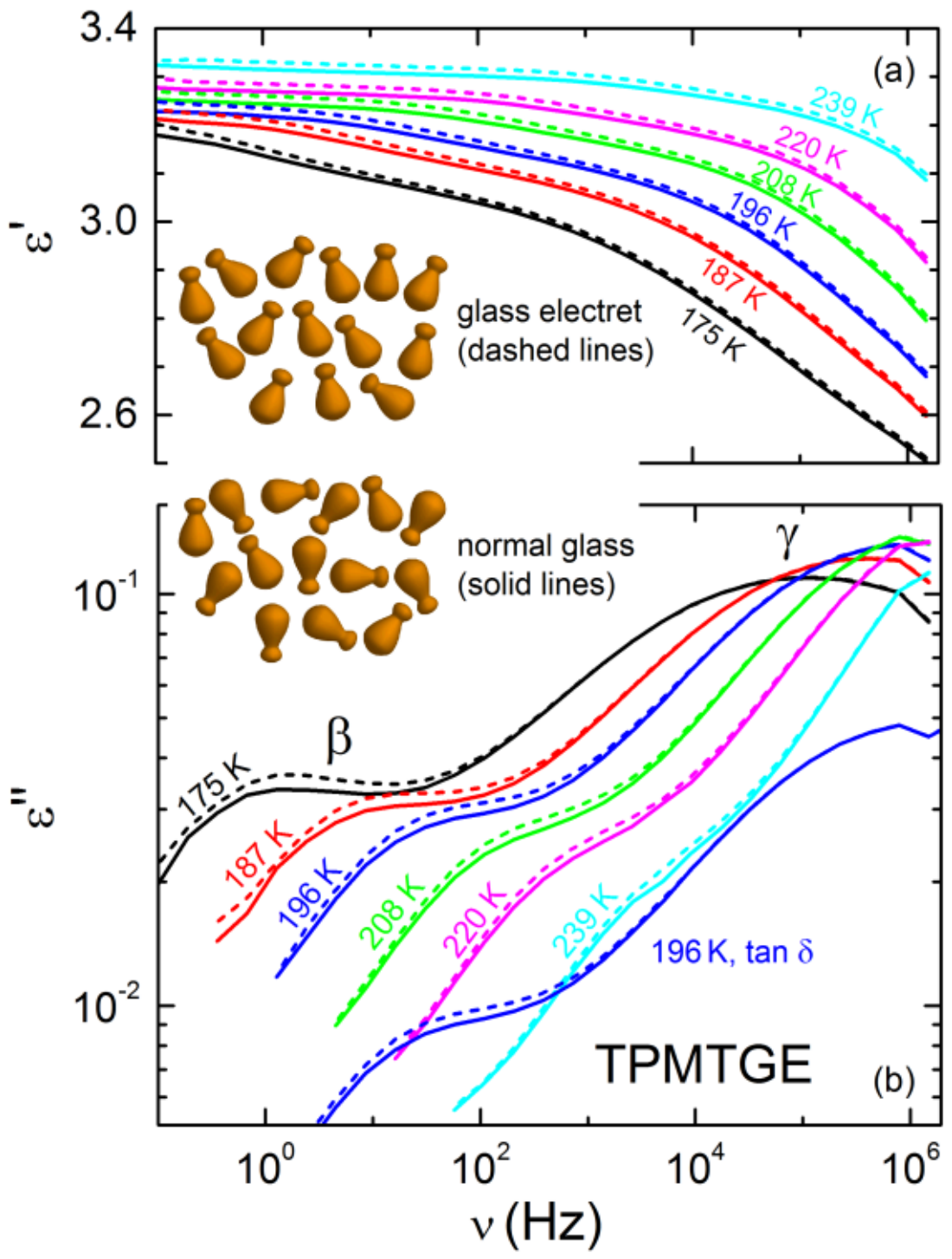

FIG. 2. Spectra of $\varepsilon'$ (a) and $\varepsilon''$ (b) as measured in the normal-glass (solid lines) and the glass-electret state (dashed) of TPMTGE in the regime of the $\beta$ and $\gamma$ relaxations. For comparison, the lower two curves for 196 K in frame (b) show the corresponding loss-tangent spectra. The inset schematically indicates the molecular arrangements in both glass states.

Figure 2 shows the $\varepsilon'$ and $\varepsilon''$ spectra in the $\beta$- and $\gamma$-relaxation region of TPMTGE, measured in the normal-glass (solid lines) and glass-electret states (dashed lines). For the latter, we find a significant enhancement of both quantities, which is most pronounced for the $\beta$ relaxation and practically absent for the $\gamma$ relaxation. Such an increase of $\varepsilon'$ and $\varepsilon''$ is also detected for DGEBA (Fig. 3), however, here both relaxations seem to be affected [42]. This becomes especially obvious in $\varepsilon'$ [Fig. 3(a)]. The inset of Fig. 3, showing the ratio of $\varepsilon''$ in both glass states for 195 K, demonstrates that even for DGEBA the electret effect is strongest in the $\beta$-relaxation regime around 100 Hz. Corresponding behavior is also found in spectra of the loss tangent, $\tan\delta = \varepsilon''/\varepsilon'$ as indicated, e.g., by the lower curves for 196 and 195 K in Figs. 2(b) and 3(b), respectively (see Fig. S3 in Supplemental Material for more temperatures [40]). As this quantity is independent of the capacitor geometry, this finding excludes that the observed effects are simply caused by a reduction of the capacitor thickness due to electrostriction. It should be noted that an enhancement of $\varepsilon''$ was also found for the $\beta$ relaxation of sorbitol when performing dielectric measurements with high *ac* fields [2,43]. In contrast, the present results were obtained with low ac field after applying a high dc field during cooling and, thus, probably have no relation to the effect observed in Refs. [2,43].

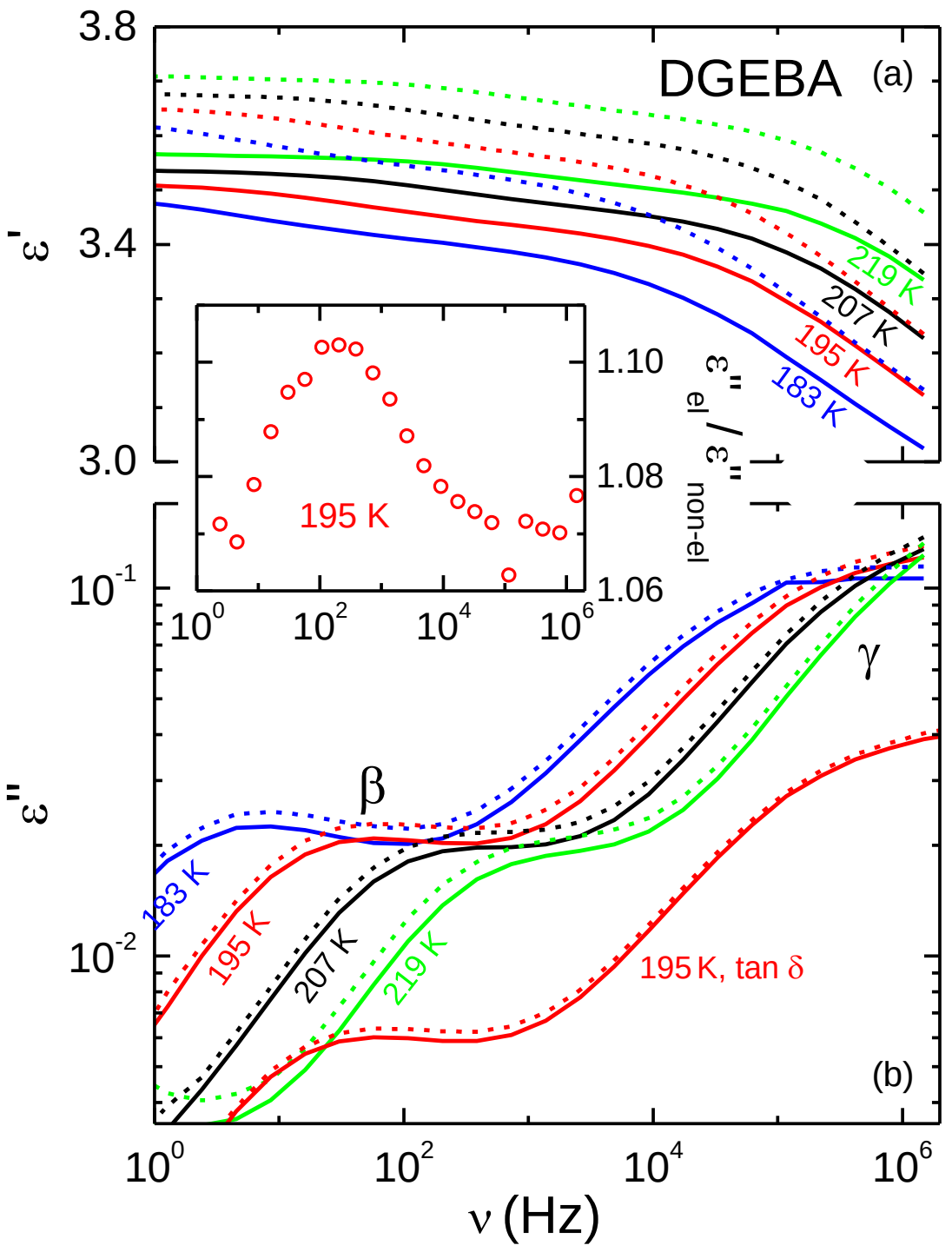

FIG. 3. Same as Fig. 2 but for DGEBA. The inset shows the ratio of $\varepsilon''$ in the glass-electret and normal-glass state at 195 K.

Now the question arises, which parameters of the relaxations have changed when comparing both glass states: relaxation time, relaxation strength, width, or several of them? To resolve this issue, we have fitted the spectra using the sum of two Cole-Cole (CC) functions [44], which are commonly employed for an empirical description of secondary relaxations [34,45,46]. The CC formula is given by

$$\varepsilon^* = \varepsilon_\infty + \frac{\varepsilon_s - \varepsilon_\infty}{1 + (i\omega\tau)^{1-\alpha}} \qquad (1)$$

where $\varepsilon^* = \varepsilon' - i\varepsilon''$ is the complex permittivity, $\varepsilon_s$ the static dielectric constant, $\varepsilon_\infty$ its high-frequency limit, $\tau$ the relaxation time, and $\alpha$ the width parameter ($0 \leq \alpha < 1$). Values of $\alpha > 0$ lead to a symmetric broadening of the loss peaks compared to the monodispersive Debye case.

As an illuminating example, Fig. 4(a) shows the fit results for the electret state of TPMTGE at a single temperature. As indicated in the figure legend, here only certain parameters were allowed to vary freely during the fitting procedure while the others were kept fixed to the values obtained from the analysis of the non-electret spectrum at the same temperature (Fig. S4 in Supplemental Material [40]). A reasonable agreement with the experimental data is already achieved when the relaxation strength, $\Delta\varepsilon = \varepsilon_s - \varepsilon_\infty$, of the JG $\beta$ relaxation is the only free parameter while all others are fixed to the non-electret values (solid line). Keeping instead $\Delta\varepsilon$ fixed and leaving the other parameters of the $\beta$ relaxation free, does not lead to fits of similar quality, even if both, $\tau$ and $\alpha$ are allowed to vary (dashed and dash-dotted lines in Fig. 4). For $\varepsilon'(\nu)$, which was simultaneously fitted, the differences are less obvious because it varies much less with frequency (Fig. S5 in Supplemental Material [40]).

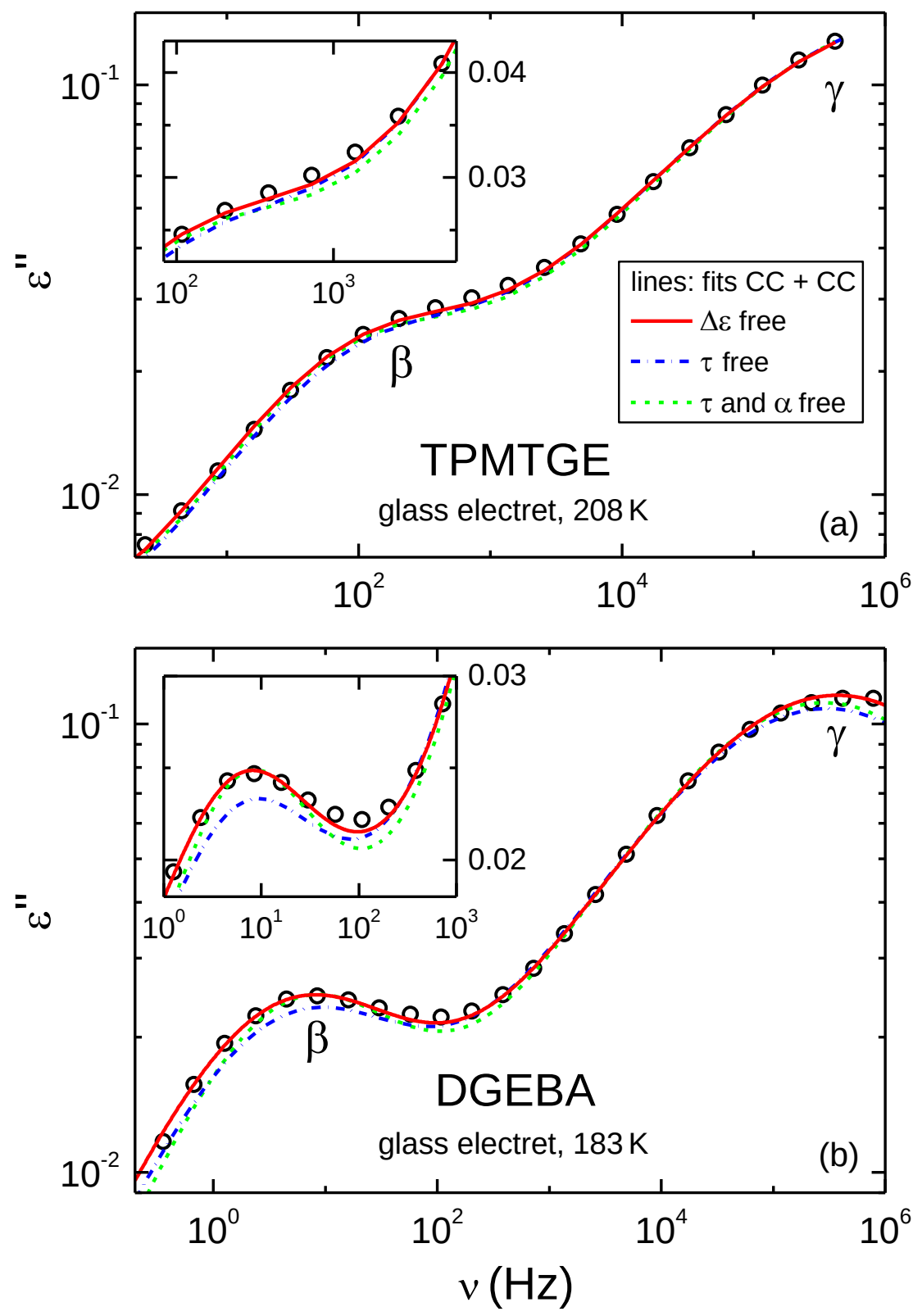

FIG. 4. Dielectric-loss spectra in the electret state of TPMTGE at 208 K (a) and of DGEBA at 183 K. The lines are fits with the sum of two CC function, eq. (1). Here part of the parameters were fixed to the values obtained for the same temperatures in the normal-glass state and only one or two parameters were allowed to vary freely as indicated in the legend of frame (a). The insets provide zoomed views on linear scales.

The results of a similar analysis for DGEBA are shown in Fig. 4(b). Here the corresponding parameters of the $\gamma$ relaxation were also allowed to be optimized during the fit, which accounts for the additional electret-induced variation in the $\gamma$-relaxation regime, documented in Fig. 3. The results are essentially comparable to those for TPMTGE: an increase of $\Delta\varepsilon$ is sufficient to explain the difference of the electret to the normal-glass spectra. While, of course, one cannot fully exclude that the other parameters may also simultaneously vary, this represents the most straightforward explanation of the experimental results. It is also in accord with a simple visual inspection of Figs. 2 and 3, leading to the impression that only $\Delta\varepsilon$ has changed in the electret state.

In summary, we have demonstrated that bringing a glass into an electret state, with partly parallel molecular orientations, significantly affects its dielectric response in the regime of the secondary relaxations. For both investigated glass formers, our findings indicate that partial molecular orientation alignment in the glass-electret state increases the relaxation strength of the JG $\beta$-relaxation. As mentioned above, preparing a glass as an electret enhances its free energy [17]. This is also true for glasses produced with a very rapid cooling rate [32]. Interestingly, for the latter case the relaxation strength of the JG relaxation, $\Delta\varepsilon_{JG}$, was also found to be higher than for glasses formed using moderate rates [27,28,29,30]. Thus, in both cases an enhancement of the free energy of a glass seems to increase its $\Delta\varepsilon_{JG}$ [47]. In Ref. [29], the higher $\Delta\varepsilon_{JG}$ of rapidly cooled glasses (and its decrease during aging) was ascribed to their more loosely packed structure leading to more (or larger) islands of mobility and, thus, more molecules participating in the JG relaxation. A glass formed by rapid cooling is farther away from equilibrium than that formed by slow cooling. One may speculate that the enforced, "unnatural" parallel orientation of the molecules in the present glass-electrets also favor the formation and growth of such regions with enhanced mobility during cooling. This would support the view that only a fraction of the total number of molecules participate in the JG relaxation. In any case, the found alteration of the dielectric properties in the glass-electret state, formed by applying a high dc field during cooling, represents a significant nonlinear dielectric effect that is different from those investigated until now (e.g., [12]). So far, non-linear dielectric spectroscopy concentrated on the primary or structural relaxation, which enabled far-reaching conclusions about the nature of glassy freezing. We hope that the present work with the focus on secondary relaxations, will stimulate further investigations of this effect, also in other classes of glass formers, necessary to reveal its general implications for the glass transition.

---

* Corresponding author.
Peter.Lunkenheimer@Physik.Uni-Augsburg.de

# Supplemental Material

for

# Johari-Goldstein relaxation in glass-electrets

P. Lunkenheimer,[1] F. Humann,[1] D. Reuter,[1] K. Geirhos,[1] A. Loidl,[1] and G. P. Johari[2]

[1]*Experimental Physics V, Center for Electronic Correlations and Magnetism, University of Augsburg, 86159 Augsburg, Germany*

[2]*Department of Materials Science and Engineering, McMaster University, Hamilton, ON L8S 4L7, Canada*

## 1. TSDC measurement

Figure S1 shows the results of a thermally-stimulated depolarization-current (TSDC) measurement of DGEBA. The detected pyrocurrent peak, occurring during heating close to $T_g \approx 255$ K, proves that the material can be *significantly* poled with a field of 133 kV/cm applied during cooling.

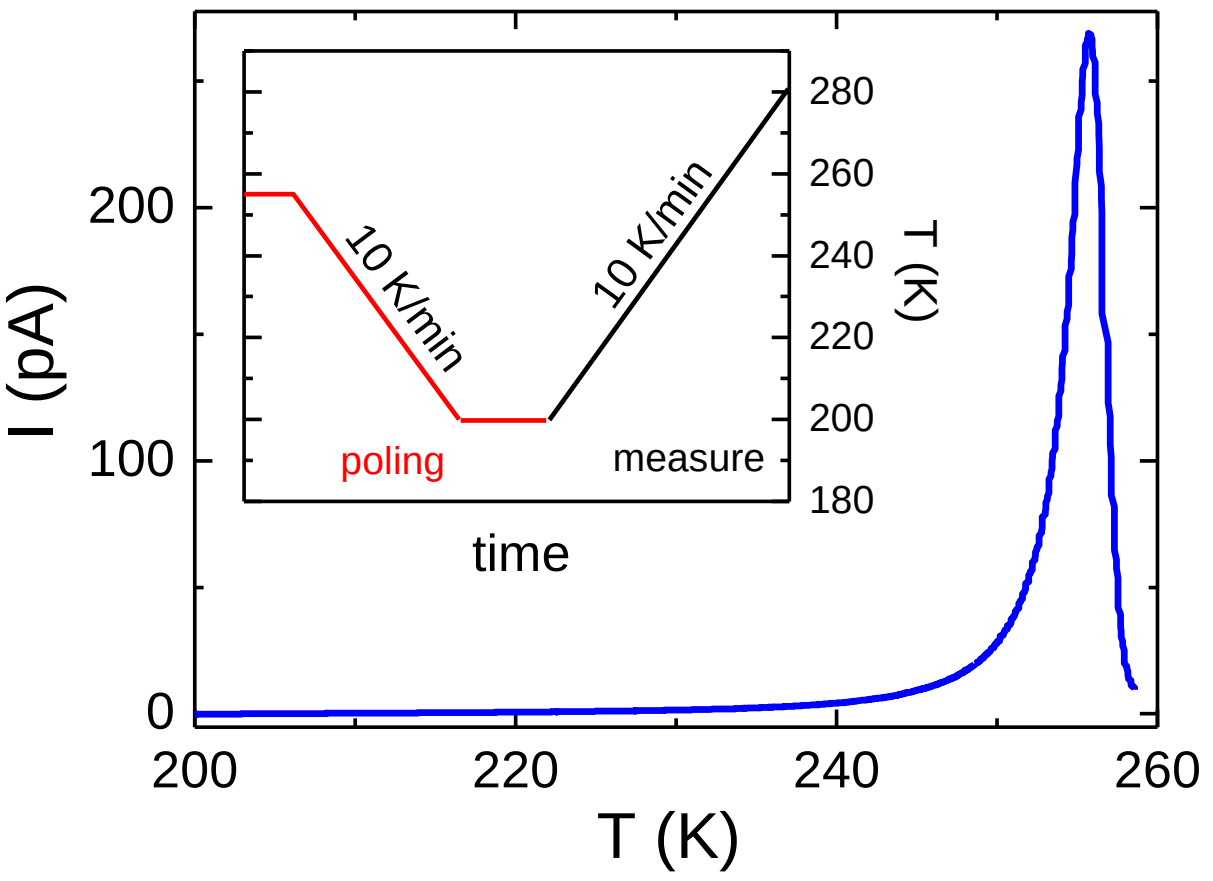

FIG. S1. Red line: Pyrocurrent as measured during a TSDC experiment under heating after the sample was cooled through $T_g$ with an applied field of 133 KV/cm. The inset schematically shows the applied temperature-time protocol.

## 2. Dielectric spectra of TPMTGE

Figure S2 shows the results of a conventional dielectric measurement of $\varepsilon'(\nu)$ and $\varepsilon''(\nu)$ of TPMTGE for various temperatures, providing an overview of the occurrence of the different relaxation processes ($\alpha$, $\beta$, and $\gamma$) in this material.

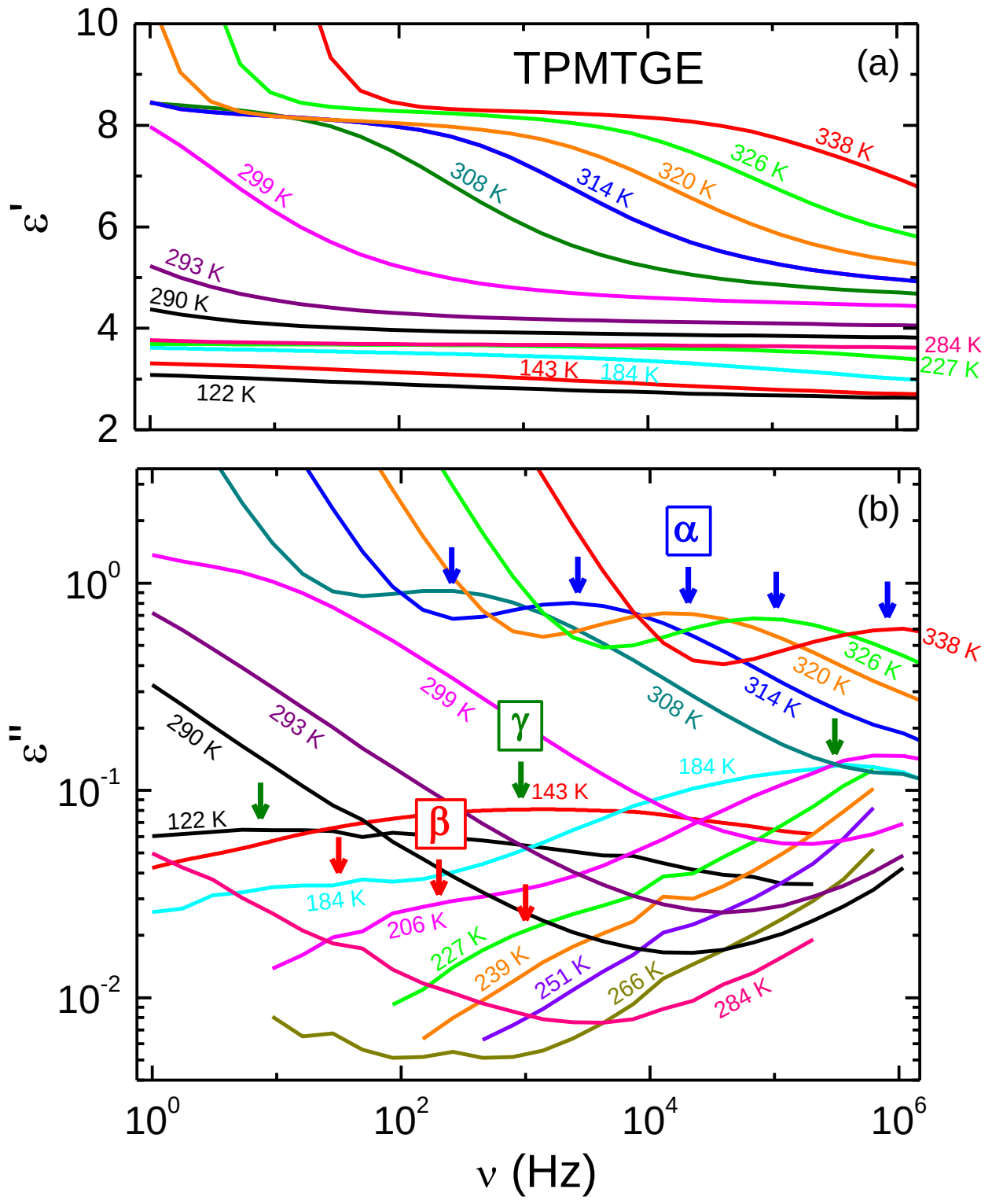

FIG. S2. Spectra of the dielectric constant $\varepsilon'$ (a) and loss $\varepsilon''$ (b) of TPMTGE at selected temperatures. The arrows indicate the peak frequencies of the three observed relaxation processes.

## 3. Loss-tangent spectra

Figure S3 shows the tan $\delta$ spectra of the normal-glass and glass-electret states of both investigated materials for various temperatures. An enhancement for the electret state is observed, which excludes a pure geometry effect, caused, e.g., by a field-induced contraction of the sample capacitor.

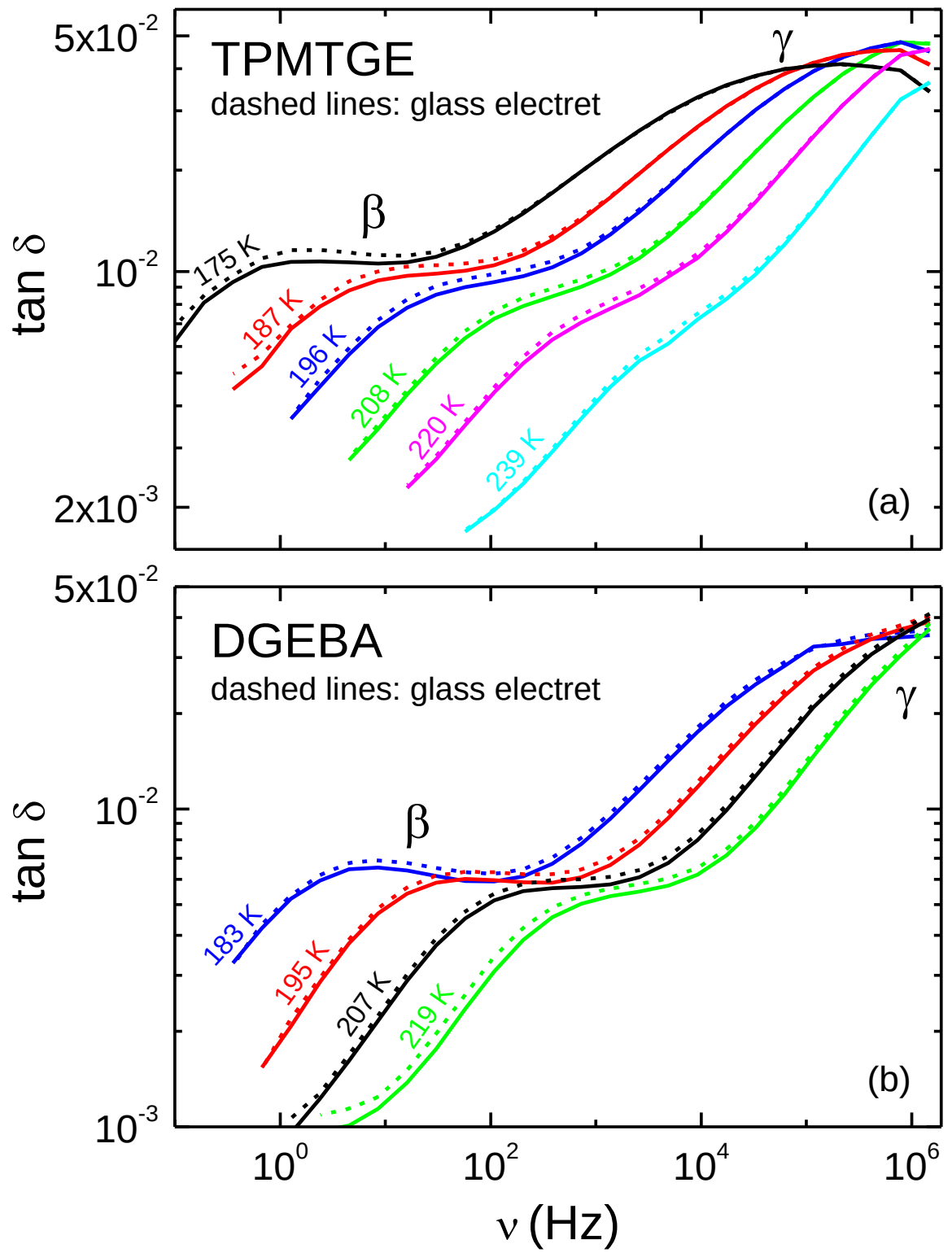

FIG. S3. Spectra of tan $\delta$ at different temperatures as measured in the normal-glass (solid lines) and the glass-electret state (dashed) of TPMTGE (a) and DGEBA (b) in the regime of the $\beta$ and $\gamma$ relaxations.

## 4. Fits for the normal glass

Figure S4 shows examples of fits of the normal-glass states of both investigated materials using the sum of two Cole-Cole (CC) functions.

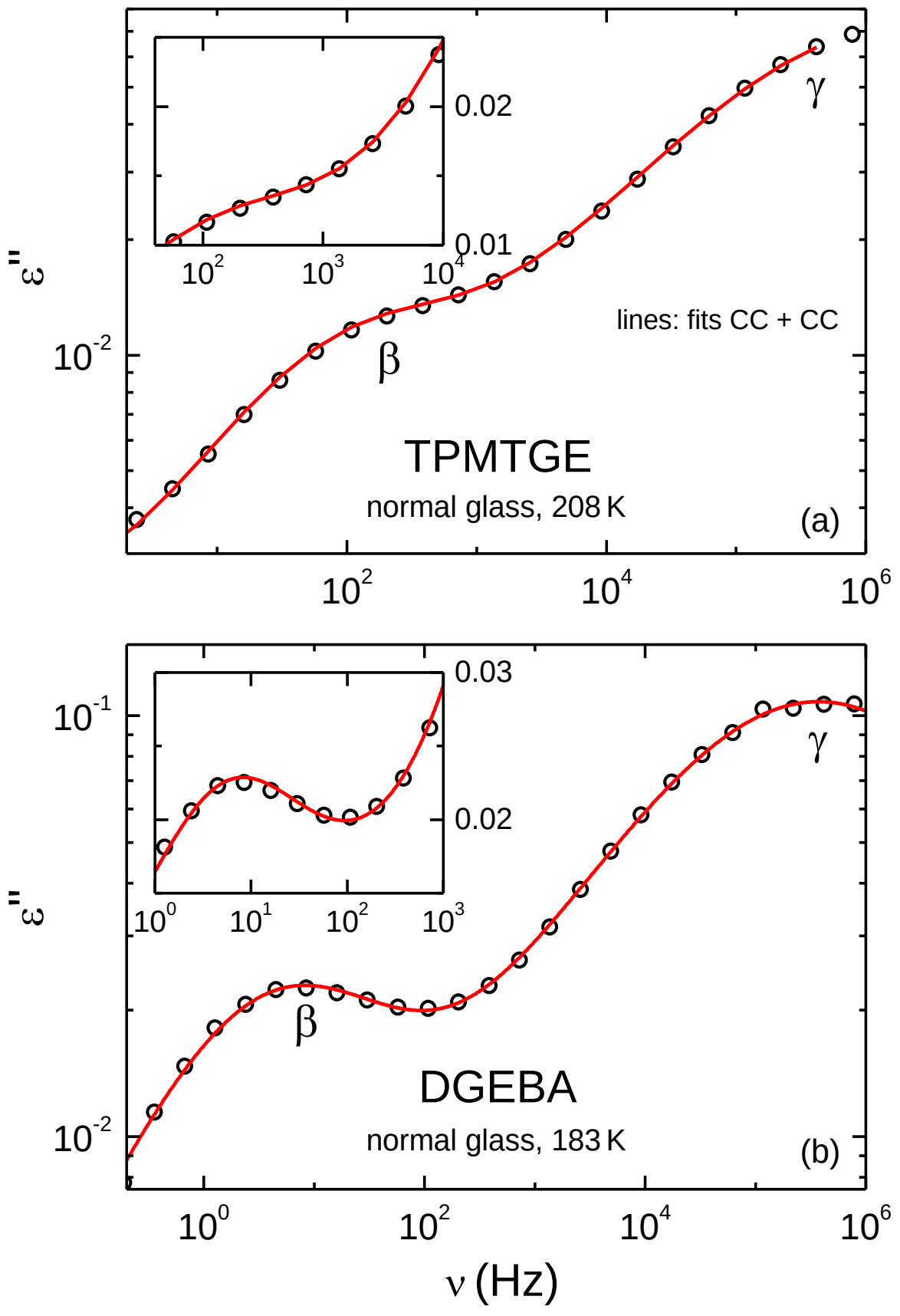

FIG. S4. $\varepsilon''$ spectra in the normal-glass state for TPMTGE at 208 K (a) and DGEBA at 183 K (b). The lines are fits with the sum of two CC functions. The inset show zoomed views.

## 5. Fits of ε' in the glass-electret

Figure S5 shows examples of $\varepsilon'(\nu)$ in the glass-electret state of both investigated materials at a single temperature. The lines are alternative fits with the sum of two CC functions. Here only certain parameters (as indicated) were allowed to vary freely during the fitting procedure while the others were kept fixed to the values obtained from the analysis of the non-electret spectrum at the same temperature. These fits were performed simultaneously with those of $\varepsilon''(\nu)$ shown in Fig. 4 of the main paper.

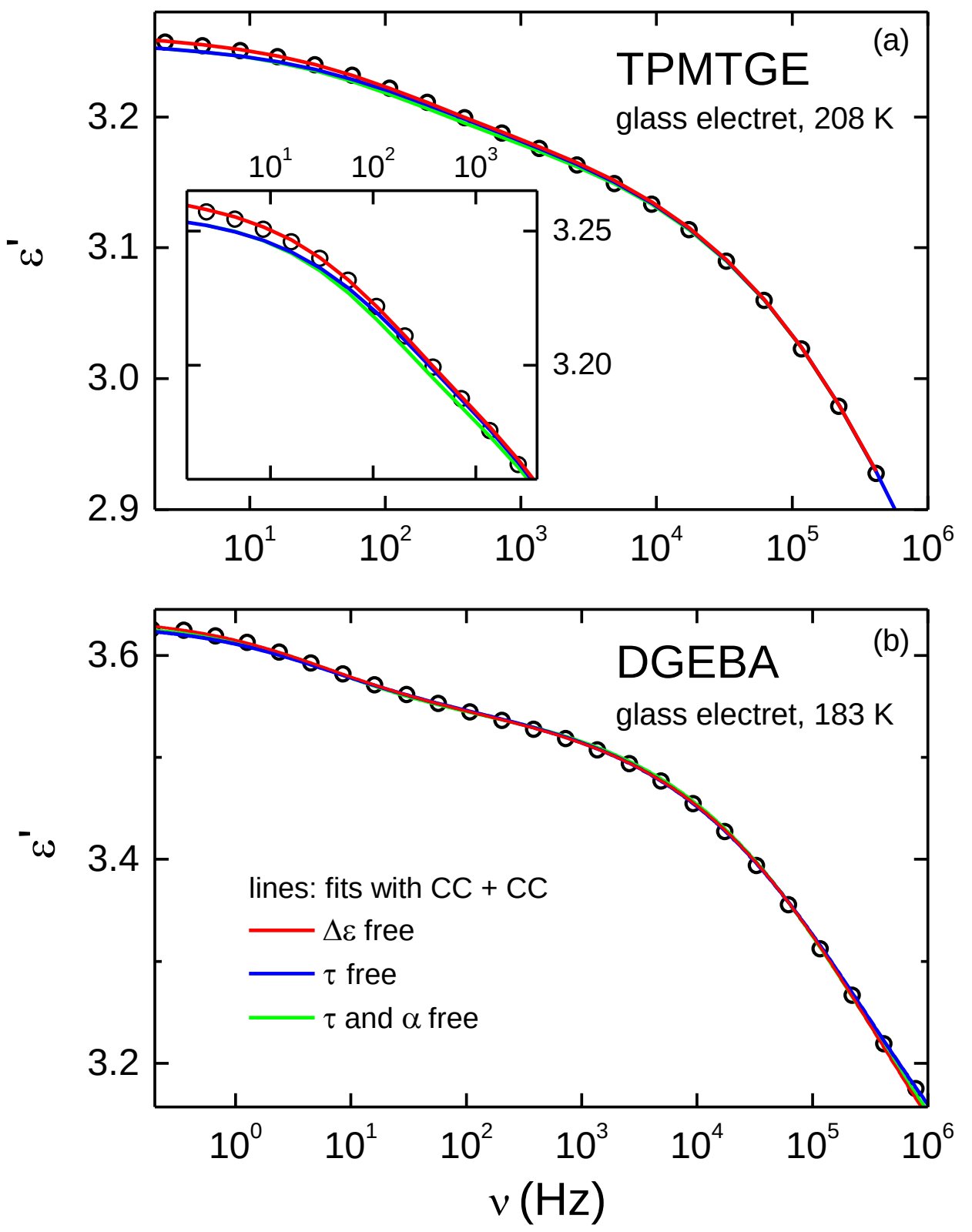

FIG. S5. $\varepsilon'$ spectra in the electret state of TPMTGE at 208 K (a) and of DGEBA at 183 K (b). The lines are alternative fits performed simultaneously for $\varepsilon'$ and $\varepsilon''$ as shown for the loss in Fig. 4 of the main paper.